\documentstyle[aps,epsf,epsfig]{revtex}
\title{Entangling Atoms and Ions in Dissipative Environments}

\author{A. Beige, S. Bose$^\dag$, D. Braun$^{\dagger\dagger}$,
S. F. Huelga$^*$, P. L. Knight, M. B. Plenio, and V. Vedral$^\dag$}

\address{Optics Section, Blackett Laboratory, Imperial College London,
London SW7 2BZ, England. \\
$^\dag$Clarendon Laboratory, University of Oxford, England. \\
$^{\dagger\dagger}$FB7, Universit{\"a}t Essen, Germany. \\
$^*$Department of Physics and Astronomy, University of Hertfordshire, Hatfield, England.}

\date{\today}

\begin{document}

\maketitle
\draft

\begin{abstract}
Quantum information processing rests on our ability to
manipulate quantum superpositions through coherent unitary transformations,
and to establish entanglement between constituent quantum components of the
processor. The quantum information processor (a linear ion trap, or a cavity
confining the radiation field for example) exists in a dissipative
environment. We discuss ways in which entanglement can be established
within such dissipative environments. We can even make use of a strong
interaction of the system with its environment to produce entanglement in a
controlled way.
\end{abstract}

\vspace*{0.2cm}

\section{Introduction}

In quantum information theory, information is
stored in the states of a quantum mechanical system instead of in the bits
of a classical computer. One possible system to realise a single quantum
mechanical bit (qubit) is given  by a two--level atom with the states
$|0\rangle$ and $|1\rangle$. The atoms can be stored for instance in a
linear trap, an optical lattice and/or inside an optical cavity. The main
advantage of using quantum mechanical systems arises from the fact that the
system can be prepared in a superposition of states. But to do so, one has
to be able to prepare the system in any arbitrary initial state. These
states are in general entangled. In addition, one has to find ways to
manipulate the state of the system in a controlled way.

The main problem in the realisation of quantum computing or quantum
information processing is decoherence \cite{Decoherence,Palma}. The reason for this is that it is
difficult to isolate a quantum mechanical system from its environment. For
instance in the case of two--level atoms, spontaneous emission leads to the
loss of information. Therefore, decoherence is usually thought of as an
entirely destructive effect. Quantum error correction codes have been
invented \cite{Steane96} and the use of decoherence-free subspaces \cite{Palma,Zanardi97,Lidar98,Guo98} for
quantum computing has been proposed.

Here we go a step further. In contrast to the widely held
folk-belief of decay being something entirely negative, we demonstrate how decay can
lead to entanglement of quantum systems rather than destroying
entanglement. Indeed, the detection of decay (or no decay) can be used as a method of state preparation, and can allow a wide range of quantum computation and
communication tasks \cite{cab,zwei,DFS,bose,DFC}.
In this paper we give examples in which
the interaction between a system and its environment is used to create
entanglement of a system, rather than destroying it. In Section II and III
we discuss the preparation and manipulation of entangled states of atoms
inside a cavity. In Section IV we show how quantum communication protocols
such as teleportation can be implemented by the detection of photon decay
from cavities containing trapped ions or atoms.

\section{Entangling two atoms inside a cavity}

First we give an example of how two atoms can be prepared in a
maximally entangled state by making use of the interaction of the system
with its environment. We will assume a strong coupling between the system
and the surrounding free radiation field. To prepare the system in a pure
state
we make use of the time evolution under the condition that no photon emissions are observed.
The scheme is very simple and should be easy to implement with present
technology. Its success rate is about 50$\,\%$. Using similar ideas, but exploiting the adiabatic elimination of fast variables in a decaying system, even higher success can be achieved. This will be discussed in the following
Section.

\subsection{The physical system}

Our system consists of two identical two-level atoms/ions confined in a linear trap or an optical lattice as shown in Fig.~\ref{fig1}. The atoms are surrounded by an optical cavity. The energy eigenstates of the atoms are denoted by $|0 \rangle_i$ and $|1 \rangle_i$ $(i=1,2)$ in the following. We assume that the distance
between the atoms is much larger than an optical wavelength which allows us
to assume that each atom can be individually addressed with a laser
pulse. The atomic transition is assumed to be in resonance with the cavity
mode and its frequency $\omega_0$ equals the frequency of the resonator mode
$\omega_{\rm cav}$. Each atom couples to the single field mode inside the
cavity with a coupling constant $g_i$. For simplicity we assume $g_1 =g_2
\equiv g$. The relaxation of the ion-cavity system can take place through
two different
channels. Each atom can spontaneously emit a photon with a rate $
\Gamma$. In addition, a photon inside the cavity mode leaks out through the
cavity mirrors with a rate $\kappa$. Note the presence of a single photon
detector $D$ in our scheme. This set up will allow us to monitor photons
leaking through the cavity mirrors. The detection efficiency varies with the
wavelength but it can be up to $90 \%$.

\begin{minipage}{6.54truein}
\begin{figure}[htb]
\begin{center}
\epsfxsize6cm
\centerline{\epsfbox{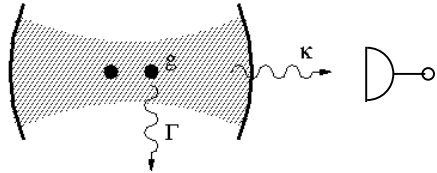}}
\caption{Experimental set-up. The system consists of two two-level
atoms placed inside a leaky cavity. The decay rate $\Gamma$ describes the
spontaneous emission of each atom, while the rate $\kappa$ refers to photons
leaking through the cavity mirrors. The latter can be monitored by a detector.}\label{fig1}
\end{center}
\end{figure}
\end{minipage}

The field annihilation operator for the cavity mode is denoted by $b$ in the following. In addition the atoms are weakly coupled to the free radiation field outside the cavity with a coupling constant $g_{{\bf k} \lambda}^{(i)}$ for the $i$th atom and a field mode with wave vector ${\bf k}$ and polarisation $\lambda$. The annihilation operator for this mode is $a_{{\bf k} \lambda}$. This free radiation field provides a ``heat bath'' for the atoms and is responsible for spontaneous emission. We take also into account non--ideal cavity mirrors by coupling the field inside the resonator to the outside with a strength $s_{{\bf k} \lambda}$, so that single photons can leak out. The annihilation operator of the free radiation field to which the cavity field couples is given by $\tilde{a}_{{\bf k} \lambda}$. Then, in the Schr\"odinger picture the Hamiltonian of the system and its environment is given by
\begin{eqnarray} \label{21}
H &=& \sum_{i=1,2} \hbar \omega_0 \, \sigma_i^\dagger \sigma_i
+ \hbar \omega_0 \, b^\dagger b
+ \sum_{{\bf k} \lambda} \hbar \omega_k
\left( a_{{\bf k} \lambda}^\dagger a_{{\bf k} \lambda}
+ \tilde{a}_{{\bf k} \lambda}^\dagger \tilde{a}_{{\bf k} \lambda} \right)
\nonumber \\
& & + {\rm i} \hbar \sum_{i=1,2} g \, b \sigma_i^\dagger
+ {\rm i} \hbar \sum_{i=1,2} \sum_{{\bf k} \lambda}
g_{{\bf k} \lambda}^{(i)} \, a_{{\bf k} \lambda} \sigma_i^\dagger
+ {\rm i} \hbar \sum_{{\bf k} \lambda} s_{{\bf k} \lambda} \,
\tilde{a}_{{\bf k} \lambda} b^\dagger + {\rm h.c.}
\end{eqnarray}
The first four terms give the interaction free Hamiltonian and correspond to the free energy of the atoms, the resonant cavity mode, and the electromagnetic fields outside the system. Going over to the interaction picture with respect to the interaction free Hamiltonian gives rise to the interaction Hamiltonian
\begin{eqnarray} \label{HI}
H_{\rm I} &=& {\rm i} \hbar \sum_{i=1,2} g \, b \sigma_i^\dagger
+ {\rm i} \hbar \sum_{i=1,2} \sum_{{\bf k} \lambda}
g_{{\bf k} \lambda}^{(i)} \, a_{{\bf k} \lambda} \sigma_i^\dagger
\, {\rm e}^{{\rm i} (\omega_0-\omega_k) t}
+ {\rm i} \hbar \sum_{{\bf k} \lambda}
s_{{\bf k} \lambda} \, \tilde{a}_{{\bf k} \lambda} b^\dagger
\, {\rm e}^{{\rm i} (\omega_0-\omega_k) t} + {\rm h.c.}
\end{eqnarray}
The first term contains the coupling of the atoms to the cavity mode. The
second term describes the coupling of the atoms to the free radiation field
and is responsible for spontaneous emission. The last term describes the damping
of the cavity mode by leaking of photons through the cavity mirrors.

\subsection{A preparation scheme}

In the following we assume that the parameters in the proposed experiment fulfill the condition
\begin{equation} \label{par}
\Gamma \ll \kappa, ~g^2/\kappa  ~.
\end{equation}
Using this parameter regime a maximally entangled state of the atoms can be
prepared in the following way. We assume that both atoms are
initially in the ground state while the cavity mode is empty. Then atom $1$ is driven into the excited state by
application of a laser pulse with the length of a $\pi$ pulse. In this way
we introduce an excitation into the system and the initial conditions for
our
scheme will be given by the composite state
\begin{equation} \label{initial}
|\psi_0\rangle
= |0\rangle_{\rm cav} \otimes |1\rangle_1 \otimes |0\rangle_2 \equiv |010\rangle~.
\end{equation}
After this $\pi$ pulse, during a short waiting time
\begin{equation} \label{dt}
\Delta t \gg 1/\kappa,\, \kappa/g^2
\end{equation}
the system may emit a photon. Whether a photon emission takes
place or not has to be measured by the detector $D$ shown in
Fig.~\ref{fig1}.
Depending on the outcome of this measurement the preparation scheme
succeeds or fails and has to be repeated again.

If the system emits a photon its excitation is lost and the state vector
is restored to the ground state $|000\rangle$, the initial state of
the scheme. But there is also the possibility that the excitation remains
inside the system. The probability for this to happen equals 50$\,\%$ and
coincides with the success rate of the scheme we propose. In this case the
atoms are prepared in the state
\begin{equation} \label{0a}
|\psi \rangle = |0a\rangle ~~{\rm with}~~
|a \rangle \equiv (|0\rangle_1 |1\rangle_2 - |1\rangle_1
|0\rangle_2)/\sqrt{2}~,
\end{equation}
which is a maximally entangled state of the two atoms. We will show later
that the prepared state $|0a\rangle$ is stable and does not change in time
as long as no additional interaction is applied.

\subsection{An intuitive explanation why the scheme works}

First we give simple arguments to show why the preparation scheme
works.  The main underlying reason is given by the fact that the atomic
states $|0\rangle_1|0\rangle_2$ and $|a \rangle$ are {\em trapped} states
\cite{zwei,knight,yeoman,meystre}. If the cavity mode is empty no
interaction between the mode and the atoms takes place, which can be proven
easily with the help of the first term on the right hand side of Eq.~(\ref{HI}). Applying this operator from Eq.~(\ref{HI}) onto a state of the general form
\begin{equation} \label{dfstate}
|\psi \rangle = \alpha \, |000 \rangle + \beta \, |0a \rangle
\end{equation}
always gives zero. Even if the atoms are excited, they cannot transfer
their excitation into the cavity. No photon leaks out through the
cavity mirrors. As a consequence, a system once prepared in a state with
an empty cavity field and a trapped state of the atoms does not change in
time. The states (\ref{dfstate}) are the {\em decoherence-free} states of the system \cite{DFS,DFC}. On the other side, if the system is prepared in a state with no overlap with $|\psi \rangle$ given in Eq.~(\ref{dfstate}) it will
emit a photon during the time $\Delta t$ for sure. The observation of the system over a time interval $\Delta t$ can be interpreted as a measurement of whether the system is decoherence--free or not.

Making now use of the projection postulate \cite{lueders} the measurement theory
suggests the following: If the initial state of the system is given by $|\psi_0 \rangle =
|010\rangle$ the system is at the end of the measurement in case of no
photon emission given by $|0a \rangle$. The probability for no photon
emission (which coincides with the success rate of the scheme) equals the
overlap of $|\psi \rangle$ with the decoherence--free subspace given by
Eq.~(\ref{dfstate}),
\begin{equation} \label{1/2}
P_0(\Delta t,\psi_0) = |\langle 000|\psi_0 \rangle|^2 + |\langle 0a|\psi_0 \rangle|^2 = {1\over 2} ~.
\end{equation}
If the initial state $|\psi_0 \rangle$ of the system is a product state with no population in state $|000\rangle$. this factor of one half is the highest success rate in preparing a maximally entangled state of both atoms which can be obtained by the scheme proposed here.

Summarising, we need the interaction between the atoms and the cavity mode to be assured that the prepared state is an entangled state. The non--Hermitian term in the no--jump evolution describing the decay of unstable states was used to remove the unwanted states during the no--photon time evolution. To make use of the no--photon case is, in general, essential because the system is otherwise prepared in a statistical mixture of states.

\subsection{Results of a detailed analysis}

The time evolution of the system under the condition of no photon emissions can easily be described using the quantum jump approach \cite{qja1,qja2}, with the help of the relevant conditional Hamiltonian $H_{\rm cond}$. This Hamiltonian is non--Hermitian and leads to a non--unitary time evolution operator $U_{\rm cond}(t,0)$. The norm of a state vector developing through $U_{\rm cond}$,
\begin{eqnarray} \label{psi0}
|\psi^0(t) \rangle &=& U_{\rm cond}(t,0) \, |\psi_0(t) \rangle~,
\end{eqnarray}
decreases in time and can be used to determine the probability to find no
photons in $(0,t)$, if the system was initially prepared in the state $|\psi_0
\rangle$,
\begin{eqnarray} \label{P0}
P_0(t,\psi_0) &=& \| \,  U_{\rm cond}(t,0) \, |\psi_0 \rangle \, \|^2~.
\end{eqnarray}
The state of the system under the condition of no photon emission at time
$t$ equals the normalised state of Eq.~(\ref{psi0}) which still has to be
normalised. The conditional Hamiltonian can be derived from the Hamiltonian $H_{\rm I}$ of Eq.~(\ref{HI}) \cite{qja1}. In the case of the system considered here we find \cite{zwei}
\begin{eqnarray} \label{28}
H_{\rm cond} &=& {\rm i} \hbar \, g \sum_{i=1,2}
\left( b \sigma_i^\dagger - {\rm h.c.} \right)
- {\rm i} \hbar \, \Gamma \sum_{i=1,2}  \sigma_i^\dagger \sigma_i
- {\rm i} \hbar \kappa \, b^\dagger b ~.
\end{eqnarray}
The first term describes as above the interaction between the atoms and the cavity mode. The last two terms lead to a decrease of the norm of a state vector if photons can be emitted or leak out through the cavity mirrors.

\begin{figure}[htb]
\begin{center}
\epsfxsize8.0cm
\centerline{\epsfbox{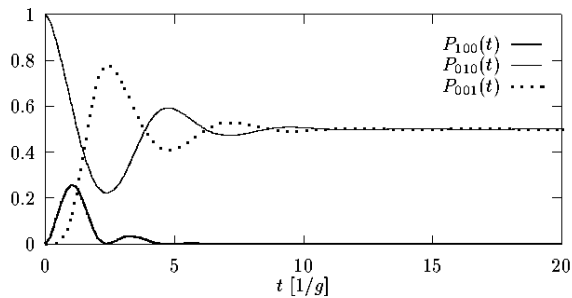}}
\caption{The time dependence of the probability amplitudes for the basis
states $|100\rangle$, $|010\rangle$ and $|001\rangle$ under the conditional
time evolution that no photon has been detected at all.
We have chosen $g=\kappa$ and $\Gamma = 10^{-3} g$. After a short time
the cavity mode has decayed and the atoms have reached the pure
entangled atomic state.}\label{fig2}
\end{center}
\end{figure}

To describe the system we only need to consider a three dimensional subspace formed by the states $|001\rangle$, $|010\rangle$, and $|100\rangle$, because the energy in the system does no change here during the conditional time evolution. In this subspace the conditional time evolution operator can be written as
\begin{equation}
U_{\rm cond}(t,0) = \sum_{j=1,2,3} {\rm e}^{-{\rm i} \lambda_j t/\hbar}
\, |\lambda_j\rangle\langle \lambda^j| ~,
\end{equation}
where $\lambda_j$ are the eigenstates of $H_{\rm cond}$, $|\lambda_j\rangle$ the corresponding eigenstates and the $\langle \lambda^j|$ are defined by the relation $\langle \lambda^j|\lambda_k \rangle=\delta_{jk}$ $(j,k=1,2,3)$. Due to the existence of the trapped state $|0a\rangle$ one of the eigenvalues, let us say $\lambda_1$, equals zero and we have $|\lambda_1\rangle = |0a\rangle$ and $\langle \lambda^1| = \langle 0a |$. The other two eigenvalues are given by \cite{zwei}
\begin{equation}
\lambda_{2/3} = {\hbar \over 2{\rm i}} \left( \kappa + \Gamma
\pm {\rm i} \sqrt{8g^2 + (\kappa-\Gamma)^2} \right)~.
\end{equation}
Using this and Eq.~(\ref{par}) and (\ref{dt}) leads to
\begin{equation} \label{prompt}
U_{\rm cond}(\Delta t,0) = |0a\rangle\langle 0a| ~,
\end{equation}
which is why the conditions on the parameters have been chosen.
With the help of Eq.~(\ref{prompt}) it can now be shown that the scheme works as predicted in Sections II.B. Fig.~\ref{fig2} shows the time evolution of the system under the condition of no photon emission as a function of time assuming the initial state $|\psi_0 \rangle$ of Eq.~(\ref{initial}), and has been obtained from a numerical solution of the Eq.~(\ref{psi0}). Already after a short time $15/g$ the system has reached the decoherence--free state $|0a\rangle$ and does not subsequently change in time.

\section{The decoherence--free states of $N$ atoms and their manipulation}

Decoherence--free states, which are in general entangled states and stable in time as long as no additional interaction is applied, do also exist, if there are more then two atoms inside the cavity. We show in the following that
by generalising the scheme discussed in the previous Section to the case of
$N$ atoms decoherence--free subspaces of much higher dimensions can be found. In addition we introduce a new scheme based on an environment induced quantum Zeno effect \cite{misra} to manipulate the states {\em inside} the decoherence--free subspace \cite{DFS,DFC,Lidar,others}. The success rate of this new scheme can, at least in principle, be arbitrarily close to 1.

\subsection{The physical system}

\begin{figure}[htb]
\begin{center}
\epsfxsize8.0cm
\centerline{\epsfbox{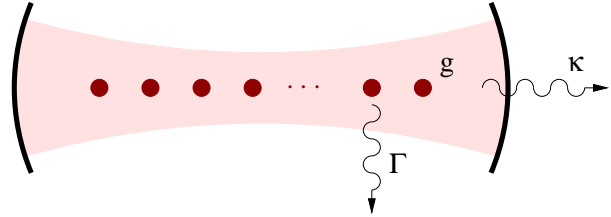}}
\caption{Schematic view of the system. The two--level atoms are held at fixed
positions in the cavity sufficiently far apart that they can be
addressed individually by laser beams.}\label{fig5}
\end{center}
\end{figure}

Now we assume that the system consists of $N$ two--level atoms inside a
cavity as shown in Fig.~\ref{fig5}. Again the atoms are assumed to be in
resonance with the single field mode. For simplicity we assume that the
coupling of all atoms to the mode is the same and we use the same notation
as in the previous Section. It will become clear below that there is no
necessity for actual detectors which measure the outcoming photons.

By analogy with Eq.~(\ref{HI}) the Hamiltonian of this system is given by
\begin{eqnarray}
H_{\rm I} &=& {\rm i} \hbar \sum_{i=1}^N g \, b \sigma_i^\dagger
+ {\rm i} \hbar \sum_{i=1}^N \sum_{{\bf k} \lambda}
g_{{\bf k} \lambda}^{(i)} \, a_{{\bf k} \lambda} \sigma_i^\dagger
\, {\rm e}^{{\rm i} (\omega_0-\omega_k) t}
+ {\rm i} \hbar \sum_{{\bf k} \lambda}
s_{{\bf k} \lambda} \, \tilde{a}_{{\bf k} \lambda} b^\dagger
\, {\rm e}^{{\rm i} (\omega_0-\omega_k) t} + {\rm h.c.}
\end{eqnarray}
In the same way as for two atoms, the conditional Hamiltonian can be derived from this and one finds
\begin{eqnarray} \label{HcondN}
H_{\rm cond} &=& {\rm i} \hbar \, g \sum_{i=1}^N
\left( b \sigma_i^\dagger - {\rm h.c.} \right)
- {\rm i} \hbar \, \Gamma \sum_{i=1}^N  \sigma_i^\dagger \sigma_i
- {\rm i} \hbar \kappa \, b^\dagger b ~.
\end{eqnarray}
It describes the time evolution of the system as long as no photon is
emitted. The two sources of decoherence in the system are again the spontaneous emission of a photon by
each atom with a rate $\Gamma$ and the leakage of a photon through the
cavity mirrors with a rate $\kappa$. As in the previous Section we
assume that the parameters $\Gamma$, $g$ and $\kappa$ fulfill condition (\ref{par}).

\subsection{Construction of the decoherence--free states}

With the help of the quantum jump approach a necessary and sufficient
criterion for a state $|\psi\rangle$ to be decoherence--free can easily be
found \cite{Guo98,DFS}. The probability for the emission of a photon in $(0,t)$ should be equal to zero if the system is once prepared in a decoherence free state, e.g.
\begin{eqnarray} \label{11}
P_0(t,\psi) &\equiv& 1 ~~~
\forall ~ |\psi \rangle \in {\rm DFS},~ \forall ~ t ~.
\end{eqnarray}
Because of condition (\ref{par}) we can neglect spontaneous emission $(\Gamma=0)$ in a very good approximation. In the following $|{\rm n} \rangle_{\rm cav}$ denotes the state with $n$ photons in the cavity and we define again $|{\rm n} \varphi \rangle \equiv |{\rm n}\rangle_{\rm cav} \otimes |\varphi\rangle$. To fulfill condition (\ref{11}) the probability density  $-{\rm d} P_0(t,\psi)/{\rm d}t|_{t=0}$ has to vanish. Using Eq.~(\ref{psi0}) and (\ref{P0}), this leads to $\langle \psi|(H_{\rm cond}-H_{\rm cond}^\dagger)|\psi\rangle=0$, and Eq.~(\ref{HcondN}) gives $|\psi \rangle = |0\varphi\rangle$.
Only if the cavity mode is empty no photon can leak out through
the cavity mirrors and the system does effectively not interact with its
environment. However, this is not yet sufficient -- the cavity mode must
{\em never} become populated. Thus, a state $|0 \varphi\rangle$ is only
decoherence--free if all matrix elements of the form $\langle {\rm n}
\varphi' | \,H_{\rm cond}\, |0 \varphi \rangle$ vanish for $n \ne 0$ and
arbitrary $\varphi'$. This is the case iff
\begin{equation} \label{DFSphi}
J_-\, |\varphi\rangle\equiv \sum_{i=1}^N\sigma_i\,|\varphi\rangle=0 ~,
\end{equation}
which defines the trapped states of $N$ atoms. Atomic states fulfilling this condition are well known in quantum optics as the Dicke states, of the form
\begin{eqnarray} \label{43}
|\varphi \rangle &=& |j,-j \rangle
\end{eqnarray}
in the usual $|j,m\rangle$ notation with $j = 0,1\,.\,.\,.\,, N/2$ or $j =
1/2,\, 3/2, \,.\,.\,.\,, N/2-1$, depending on whether $N$ is even or odd
respectively. Here $j$ is a quantum number with $n=N/2-j$, if $n$ is the
number of excitations in the system. The dimension of all possible trapped
states for a fixed number of atoms $N$ equals therefore \cite{DFS}
\begin{eqnarray}
\left( \begin{array}{c} N \\ (N+1)/2 \end{array} \right)
& ~~ {\rm or} ~~ & \left( \begin{array}{c} N \\ N/2 \end{array} \right)
\end{eqnarray}
depending on whether $N$ is odd or even. For large atom numbers the
dimension of the decoherence--free subspace equals in a good approximation
$(2/(\pi N))^{1/2} \cdot 2^N$ and thus increases almost exponentially with
the atom number $N$.

For two atoms there are two independent decoherence--free states,
$|000\rangle$ and $|0a\rangle$, as shown in the previous Section. With the
help of these two states also the trapped states of $N$ atoms in the cavity
can be constructed. One can show that each state of the form
$|\varphi \rangle = |0 \rangle_2 \otimes |a \rangle_{13} \otimes |a
\rangle_{45} \otimes \,. \,. \,. \, \otimes |0 \rangle_N$
in which for instance the second atom is in the ground state, the first and
third are in an antisymmetric state and so on fulfills condition
(\ref{DFSphi}). Writing down all possible states of this form leads to an
overcomplete basis which can then be normalised.

\subsection{Manipulation of the states}

We assume that the $N$ two--level atoms are confined to fixed positions
inside the cavity and spatially well separated so that laser pulses can be
applied
to each atom individually. We denote the complex Rabi frequencies for atom
$i$ by $\Omega_i$. In the presence of the laser field the conditional
Hamiltonian becomes the sum of $H_{\rm cond}$ given in Eq.~(\ref{HcondN})
and the laser Hamiltonian $H_{\rm Laser\, I}$,
\begin{eqnarray} \label{hlaser}
H_{\rm laser\,I} &=& {\hbar \over 2} \sum_{i=1}^N \Omega_i \,
\sigma_i + {\rm h.c.}
\end{eqnarray}
In the following we assume that the laser pulse is very weak and that all
non vanishing Rabi frequencies $\Omega_i$ fulfill the condition
\begin{eqnarray} \label{par2}
\Gamma \ll |\Omega_i| \ll \kappa,~g^2/\kappa~.
\end{eqnarray}
In this parameter regime spontaneous emission can be neglected in first
approximation and we can show that the system, once prepared in a
decoherence--free state cannot move out of the subspace. An explanation for this is given below.

In the presence of the laser field the effect of the interaction of the
system with its environment over a time interval $\Delta$ of the order of  $15/\kappa$ and $15\kappa/g^2$, can be interpreted as a measurement whether the system is in a decoherence free state or not because $\Delta t$ is, due to the conditions (\ref{dt}) and (\ref{par2}), much smaller than $1/|\Omega_i|$. The effect of the laser during a time interval $\Delta t$ can be
neglected. This underlying process has been discussed in more detail on the example for two atoms in the previous Section.

Here the system continuously interacts with its environment and continuous measurements are performed on the system. Therefor the effect of a weak laser pulse on the system can easily be interpreted with the help of the quantum Zeno effect \cite{misra}. The quantum Zeno effect is a consequence of the projection postulate \cite{lueders} and has been verified experimentally \cite{itano}. It predicts a slowing down of the time evolution of a system due to rapidly repeated ideal measurements. In the case of vanishing times between subsequent measurements the system is ``frozen" in the state or {\em subspace of states} it is once found in. Applying this to the system we discuss here, the quantum Zeno effect suggests: If the state of the system belongs initially to the decoherence--free subspace it will remain decoherence--free with a very high probability, even if the laser excites transitions out of the subspace.

But only the transitions out of the decoherence--free subspace are inhibited. The time evolution of the system inside the decoherence--free subspace is not affected and the state of the system at the end of a laser pulse of length $t$ equals in a very good approximation
\begin{eqnarray}
|\psi(t) \rangle &=&
I\!\!P_{\rm DFS} \,.\,.\,.\,  I\!\!P_{\rm DFS} \, U_{\rm cond} (\Delta t,0) \, I\!\!P_{\rm DFS} \, \, U_{\rm cond} (\Delta t,0) \,
|\psi_0 \rangle ~\equiv~ U_{\rm eff}(t,0) \, |\psi_0 \rangle ~,
\end{eqnarray}
where $I\!\!P_{\rm DFS}$ is the projector on the decoherence--free states
and $|\psi_0\rangle$ the initial (decoherence--free) state of the system. The operator $U_{\rm
eff}$ is an {\em effective} time evolution operator. Neglecting spontaneous emission $(\Gamma=0)$ leads to the effective Hamiltonian
\begin{eqnarray} \label{heff}
H_{\rm eff} &=& I\!\!P_{\rm DFS} \, H_{\rm laser\,I} \, I\!\!P_{\rm DFS}~.
\end{eqnarray}
This Hamiltonian describes the effect of the weak laser pulse to a very good approximation. Note that it is independent of the specific values of  the parameters $g$ and $\kappa$.

The states of the decoherence--free subspace can be used to obtain stable
qubits for quantum computing. In addition, the effective Hamiltonian allows
to prepare the system in an entangled state and to perform gate operations
{\em inside} the subspace. An example for a possible realisation of a nearly
decoherence--free quantum computer is given in Ref.~\cite{DFC}. In Ref.~\cite{bell} an experimental test of Bell's inequality using atoms is proposed. The scheme is based on the ideas discussed here, but due to the use of an additional level it also works with a very high efficiency if $\Gamma$ and $g$ are of the same order of magnitude.

\subsection{Example}

\begin{figure}[htb]
\begin{center}
\epsfxsize8.0cm
\centerline{\epsfbox{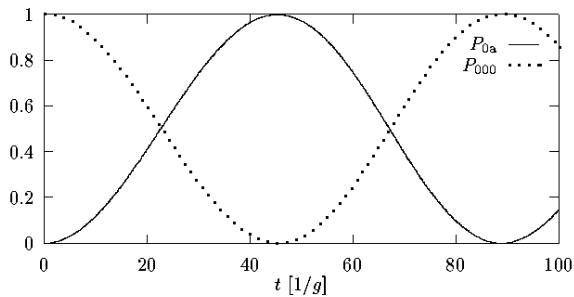}}
\caption{The time evolution of the population of the decoherence--free
states of two atoms in presence of a laser pulse focused on atom 1 derived from numerical solution of the full dynamics of Eq.~(\ref{psi0})--(\ref{28}). The
coupling constant $g$ has been chosen equal to $\kappa$ and it is
$\Omega_1=0.05 \,g$, $\Omega_2=-\Omega_1$ and
$\Gamma=10^{-3}\,g$.}\label{rot}
\end{center}
\end{figure}

To check how well the proposed scheme works, let us reconsider the case of two two--level atoms inside the cavity. In this case Eq.~(\ref{HcondN}), (\ref{hlaser}), and (\ref{heff}) lead to the effective Hamiltonian
\begin{eqnarray} \label{heff2}
H_{\rm eff} &=& {\hbar \over 2 \sqrt{2}} \, \Big[ \,
(\Omega_1-\Omega_2) \, |000 \rangle \langle 0a | + {\rm h.c.} \, \Big] ~,
\end{eqnarray}
so that to first approximation the laser just leads to oscillations between the two decoherence--free states $|000\rangle$ and $|0a\rangle$. This is in good agreement with Fig.~\ref{rot} which results from a numerical solution of Eq.~(\ref{psi0})--(\ref{28}). A more detailed analysis of the time evolution of this system  can be found in Ref.~\cite{DFS}.

\begin{center}
\begin{figure}[h]
\epsfig{file=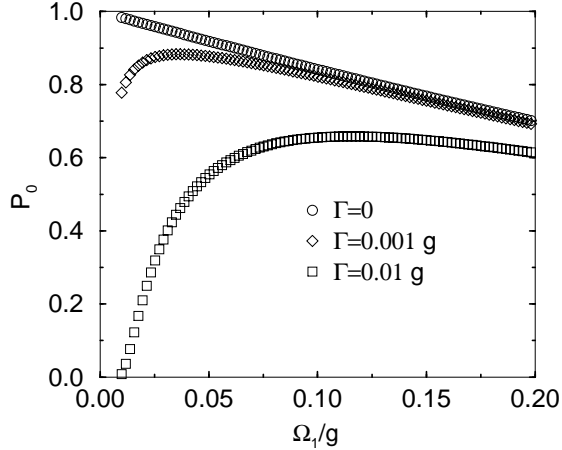,height=7cm} \\[0.2cm]
\caption{Probability for successful preparation of the maximally entangled
state $|a\rangle$ as a function of the Rabi frequency $\Omega_1$ for
$\Omega_2=-\Omega_1$, $\kappa=g$, and different values of $\Gamma$. }\label{p0}
\end{figure}
\end{center}

By choosing the length of the laser pulse in an appropriate way the scheme can also be used to prepare the atoms in the maximally entangled state $|a\rangle$.  Fig.~\ref{p0} shows the success rate $P_0$ for this scheme and also results from a numerical solution. For zero spontaneous emission, success rates arbitrarily close to unity can be achieved by reducing the Rabi frequency $\Omega_1$. However, for $\Gamma \neq 0$ this is not possible without
increasing the probability of occurance of a spontaneously emitted photon. For finite values of $\Gamma$ there is an optimal value of $\Omega_1$ for which the success rate of the preparation scheme has a maximum. The fidelity of the prepared state can, for a very wide parameter regime, be very close to 1 \cite{DFS}.

\section{Teleportation implemented by the detection of photons}

Another example in which cavity decays can play a constructive role is
in the quantum teleportation (see Ref.\cite{ben} for teleportation)
 of an atomic state. Our example derives from earlier examples of entangling distinct atoms by detection
(or the non--detection) of decays  \cite{cab,zwei}.
In all experimental implementations of teleportation to date
\cite{zei2}, the stationary
qubits have been of transient "flying" optical origin.
In earlier proposals of atomic state
teleportation  \cite{atm}, the flying qubits were
atomic states and are therefore not ideal for long distance teleportation. Our
scheme differs from these earlier in using both the ideal stationary (atomic) and
the ideal flying (photonic) qubits. It also differs crucially from the much studied
quantum communication protocols in which a photon {\em directly} transfers quantum information
from an atom trapped in a cavity to another atom in a
distant cavity \cite{zol3,Pel}.
 Our scheme does not require a
 direct carrier of quantum information between
distant atoms. Joint detection of photons {\em leaking out} of
distinct cavities enables {\em disembodied} transfer of quantum information
from an atom in one of the cavities to an atom in the other.  We thus
provide a quantum state transfer scheme that does not require us to inject
a photon into a cavity from outside \cite{zol3,Pel}. It should also be made
clear that our scheme is very different from establishing prior entanglement
between two distant atoms and subsequent teleportation of the unknown
state of a third atom using this
entanglement. In that case we would require at least two atoms trapped in
one cavity. For the basic operation of our scheme, it is sufficient to
have only one atom in each cavity. For a failure--proof operation, however,
we would require an additional atom in one of the cavities.

\begin{figure}
\begin{center}
\leavevmode
\epsfxsize=7cm
\epsfbox{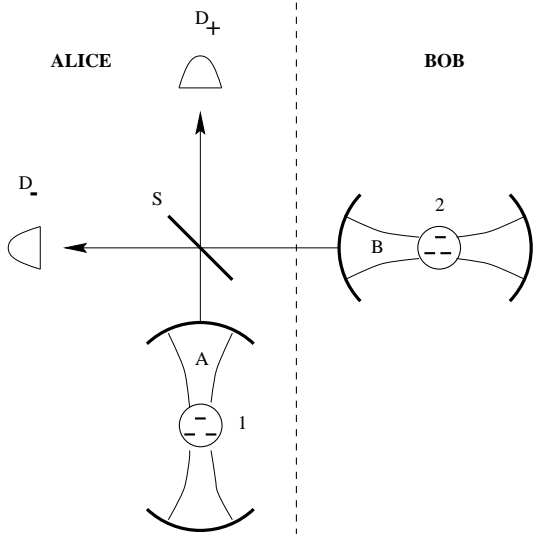}
\caption{The atomic state teleportation setup.
The cavity A, atom $1$, beam splitter $S$  and the detectors $D_+$ and $D_-$
belong to Alice, while the cavity B and atom $2$ belong to Bob.}
\label{setup}
\end{center}
\end{figure}

The setup consists of two optical
cavities, each containing a
single trapped $\Lambda$ three level atom, as shown
in Fig.\ref{setup}. Atoms $1$ and $2$ are trapped in cavities A
and B (supporting cavity modes A and B) respectively. The photons
leaking out from both the cavities are incident
on the $50-50$  beam splitter $S$ and are detected at $D_+$ and $D_-$. We assume unit efficiency
detectors initially (we include finite efficiency
later). The cavity A, atom $1$, beam splitter $S$ and the
detectors $D_+$ and $D_-$ belong to Alice. The cavity B with atom $2$ belongs
to Bob. We require both the cavities to be {\em one sided}
so that the only leakage of photons occur through the
sides of the cavities facing $S$. By following our teleportation protocol, Alice can teleport an
unknown state of her atom $1$ to the atom $2$ held by Bob in three
stages.

\vspace{-0.5cm}

In the preparation stage, Alice maps her atomic state to her cavity
state \cite{par}. At the same time Bob creates a maximally entangled state of his atom and his cavity mode. In the next stage (the detection stage) Alice waits
for a {\em finite} time for either or both of her detectors to click. If any
 one of the detectors register a single click during this time period, then
the protocol is successful. Otherwise Alice informs Bob about her failure.

 This protocol can be related to the standard teleportation protocol
\cite{ben}
by noting that the beam splitter and the
detectors constitute a device for measurement
of the joint state of the two cavities in the basis $\{|0\rangle_A
|0\rangle_B, |1\rangle_A
|1\rangle_B, \frac{1}{\sqrt{2}}(|0\rangle_A
|1\rangle_B + |1\rangle_A
|0\rangle_B), \frac{1}{\sqrt{2}}(|0\rangle_A
|1\rangle_B - |1\rangle_A
|0\rangle_B) \}$. Here $\{|0\rangle_A, |1\rangle_A \}$ and
$\{|0\rangle_B, |1\rangle_B \}$ are photon number states
in cavities A and B respectively. The teleportation is probabilistic, because it is successful
only for the pair of Bell state outcomes of the above measurement (later we
describe how to convert this to a
{\em reliable} state transfer protocol). At the end of the detection
period, if the protocol has been successful, Alice lets Bob know whether
$D_+$ or $D_-$ had clicked. This corresponds to the classical communication
part of the standard teleportation protocol \cite{ben}. Dependent on this information Bob applies a local unitary
operation to his atom to obtain the teleported state, in a
post-detection stage.

\begin{figure}
\begin{center}
\leavevmode
\epsfxsize=8cm
\epsfbox{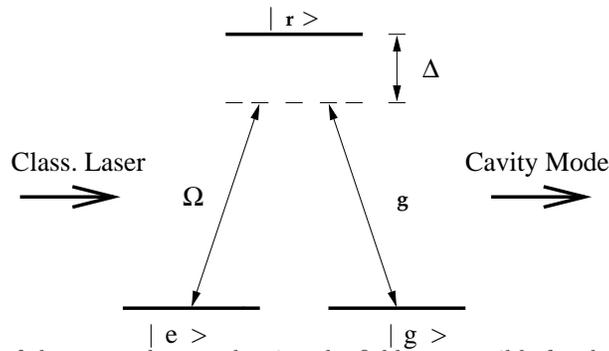}
\caption{The level configuration of the trapped atom showing
the fields responsible for the adiabatic evolution. The $|e\rangle \rightarrow |r\rangle$ transition being driven by
a classical laser field of coupling $\Omega$ and the $|r\rangle \rightarrow |g\rangle$ transition being driven by
the quantized cavity mode of coupling $g$. $\Delta$ is the detuning of both the
classical laser field and the quantized field mode from their
respective transitions.}
\label{lamdatom}
\end{center}
\end{figure}

   The detailed analysis of the scheme has been given in Ref.\cite{bose}, and
only the results will be presented
here. We looked at single realizations
conditioned on detection (or not) of cavity decays using the
quantum jump
approach \cite{qja1}. The trapped atoms are considered to be
three level atoms as shown in Fig.\ref{lamdatom}.

\vspace{-0.5cm}
Alice and Bob use two types of time evolutions of the atom-cavity system as
their basic local operations. The first type is an adiabatic evolution (shown
in Fig.\ref{lamdatom}) which is initiated by
switching on a classical laser field which
drives the $|e\rangle \rightarrow |r\rangle$ transition with a coupling
constant $\Omega$. The $|r\rangle \rightarrow |g\rangle$ transition is driven by the quantized cavity mode of coupling $g$. Both the classical laser field
and the cavity modes are assumed to be detuned from their respective
transitions by the same amount $\Delta$. As the atom is considered to be trapped at a
specific position
in the cavity, we can assume that the couplings $\Omega$ and $g$
remain constant during
the interaction.  We choose parameters such that
 $g \Omega/\Delta^2 \ll 1 $ (the upper
level $|r\rangle$ can then be decoupled from the evolution) and
$\Delta \gg \gamma$ (the spontaneous
decay rate from $|r\rangle$ can be neglected). The Hamiltonian
for the evolution of the system under such conditions (and assuming $g=\Omega$ for
simplicity), is given
by
$H^{(1)}=E |e\rangle \langle e| + E |g \rangle \langle g|+E (c |e\rangle \langle g| +  c^{\dagger} |g \rangle \langle e|)$
where $E =g \Omega/\Delta$ \cite{Pel} and $c$ and $c^{\dagger}$ are
destruction and creation operators for the cavity mode.
The other local operation
accessible to Alice and Bob is the Zeeman evolution
 used to give an arbitrary phase shift of the level $|e\rangle$ relative
to the level $|g \rangle$. The Hamiltonian for this evolution
is $H^{(2)}=\delta E |e\rangle \langle e|$, where $\delta E$ is an energy
difference. The decay rate from both cavities is assumed to be the
same and equal to $\kappa$.

 Let the unknown state of the atom $1$ which Alice wants to
teleport be
\begin{equation}
|\Psi \rangle^I_1 = a|e\rangle_1 + b|g\rangle_1 ,
\end{equation}
where the superscript $I$ in $|\Psi \rangle^I_1$ stands for input and $a$ and
$b$ are complex amplitudes. We assume that the initial state of Alice's cavity is
$|0\rangle_A$  and the initial state of Bob's atom-cavity system is
$|e\rangle_2|0\rangle_B$.
Alice first maps the state of atom $1$ onto the cavity
mode A by switching the Hamiltonian $H^{(1)}$ on for a period of time $t_I$ given by $\tan{\frac{\Omega_\kappa t_I}{2}}=-\frac
{\Omega_\kappa}{\kappa}$ where $\Omega_\kappa = \sqrt{4E^2-\kappa^2}$.
Subject to no decay being recorded in the detectors,
the cavity state is given by
\begin{equation}
|\Psi \rangle^I_A = \frac{1}{\sqrt{|a|^2\alpha^2+|b|^2}}(a \alpha |1\rangle_A + b |0\rangle_A),
\end{equation}
where $\alpha=(\frac{e^{-\frac{\kappa}{2}t_I}}{ \Omega_\kappa} 2E \sin{\frac
{\Omega_\kappa t_I}{2}})$. The probability that no photon decay takes place
during this evolution is given by $P_{ND}(A)=(|a|^2 \alpha+|b|^2)$.
Meanwhile, Bob also switches on the Hamiltonian $H^{(1)}$ in his cavity for a
shorter length of time  $t_E$ given by $\tan{\frac{\Omega_\kappa t_E}{2}}=-\frac
{\Omega_\kappa}{2E-\kappa}$. His atom-cavity system
 thus evolves to the entangled state
\begin{equation}
|\Psi \rangle^E_{2,B} =  \frac{1}{\sqrt{2}} (|e\rangle_2|0\rangle_B
+i|g\rangle_2|1\rangle_B).
\end{equation}
The probability that no photon decay takes place
during this evolution is given by $P_{ND}(B)=|\beta|^2$ where $\beta=\frac{e^{-\frac{\kappa}{2}t_E}}{\Omega_\kappa}2\sqrt{2} E \sin{\frac
{\Omega_\kappa t_E}{2}}$.
For simplicity, we assume that Alice and Bob
synchronize their actions such that the preparation of the states
$|\Psi \rangle^I_A$ and $|\Psi \rangle^E_{2,B}$ terminate at the same
instant of time.  Now comes the detection stage, in which Alice simply waits for
any one of the detectors $D_+$ or $D_-$ to click over a finite
detection time denoted by $t_D$.  Alice and Bob reject the cases in which Alice does not register any click,  or when she registers
two clicks. In the post detection
stage, Bob uses $H^{(2)}$ to give $|g\rangle_2$ an extra phase shift with
respect to $|e\rangle_2$. This phase shift is $-i$ if
$D_+$ had clicked and $i$ if $D_-$ had clicked.  This concludes the entire protocol.

 The average density matrix of Bob's atom generated through to our teleportation
procedure is given by
$\rho^{Tel}_2 = \{P_{ND}(A)
|\Psi\rangle_2 \langle \Psi|_2 + 2|a|^2\alpha^2 e^{-2 \kappa t_D} |g\rangle_2
\langle g|_2 \}/\{P_{ND}(A)+2|a|^2\alpha^2 e^{-2 \kappa t_D}\}$,
where $|\Psi\rangle_2=(a \alpha |e\rangle_2 +
b |g\rangle_2)/\sqrt{|a|^2\alpha^2+|b|^2}$.
The fidelity of this state with respect to the input state is
$F(t_D,a,b)=\{P_{ND}(A)(|a|^2\alpha+|b|^2)+2|a|^2\alpha^2 e^{-2 \kappa t_D}|b|^2\}/\{P_{ND}(A)+2|a|^2\alpha^2 e^{-2 \kappa t_D}\}$.
 We see that apart from the system parameters
$\kappa$ and $\Omega_\kappa$, the fidelity of the generated state also depends on the detection
time $t_D$ and the modulus of the amplitudes $a$ and $b$ of the initial state. It is
a teleportation protocol with a {\em state
dependent fidelity}.

  \begin{figure}
\begin{center}
\leavevmode
\epsfxsize=8cm
\epsfbox{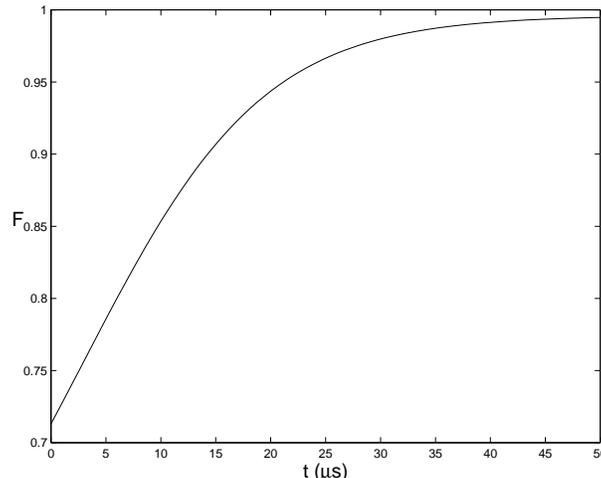}
\caption{The improvement of average teleportation fidelity with the length of the detection stage. The parameter regime is $(g:\Omega:\kappa:\gamma:\Delta)/2\pi=(10:10:0.01:1:100)$
MHz}
\label{Tfid}
\end{center}
\end{figure}

\vspace{-0.5cm}
We plot the variation of the average fidelity of teleportation over all possible input states
as a function of the detection time $t_D$ in Fig.\ref{Tfid}. We see that
the fidelity increases with increasing detection time. This
happens because increasing the detection time decreases
the proportion of $|g\rangle_2
\langle g|_2$ in the teleported state $\rho^{Tel}_2$ and brings it closer
to the initial state $|\Psi \rangle^I_1$ of Alice's atom. The parameter regime  used
for Fig.\ref{Tfid} \{$(g:\Omega:\kappa:\gamma:\Delta)/2\pi=(10:10:0.01:1:100)$
MHz\},
is carefully
chosen to satisfy all
our constraints ($g \Omega/\Delta^2 \ll 1, \Delta >> \gamma,
\Omega_\kappa >> \kappa$).  This regime could be approached, for example, by increasing the cavity finesse of Ref.\cite{kim} by an
order of magnitude and increasing the length of that cavity to
about a millimeter while
keeping the beam waist and other parameters constant.
 As evident from
Fig.\ref{Tfid}, the average fidelity exceeds $0.99$
for a detection time
of about half the cavity life time. For detectors of finite efficiency
$\eta$, the fidelity of the teleported state would be
$\{\eta P_{1D}(0,t_D) F(t_D,a,b) + 2 \eta(1-\eta) (1- P_{ND}(0,t_D)-P_{1D}(0,t_D))|b|^2\}/P_{\mbox{\scriptsize{suc}}}(\eta)$, where $P_{ND}(0,t_D)$ is
the probability of no decay during the
detection period. For a $\eta$ of $0.6$ and detection times large compared
to the cavity decay time, the fidelity of the state
in Bob's cavity becomes $\sim 0.81$.

   The above probabilistic teleportation
protocol can be modified to {\em teleportation
with insurance}, so that in the cases when the protocol is
unsuccessful, the original state of Alice's atom $1$ is not destroyed,
but mapped onto another reserve atom $r$ trapped in Alice's cavity.
To accomplish this, Alice has to follow the {\em local redundant
encoding} of Ref.\cite{van1} and codes her initial state $|\Psi \rangle^I_1$
as $a (|e\rangle_1 |g\rangle_r + |g\rangle_1 |e\rangle_r) +
b (|g\rangle_1 |g\rangle_r + |e\rangle_1 |e\rangle_r)$. After this, she
just follows the same protocol as before. But in cases when the protocol
is unsuccessful, she is left with either
the state $a |g\rangle_r + b |e\rangle_r$
or a state that can be converted to $a |g\rangle_r + b |e\rangle_r$ by a
known unitary transformation. She can now exchange the roles of atom $1$
and atom $r$ and try to teleport the state $|\Psi \rangle^I_1$ again.
She can repeat this procedure until teleportation
is successful (Of course, this holds true perfectly only when $\eta=1$).

  In this section, we have shown  how the state of an atom trapped in a cavity can be teleported to
a second atom trapped in a distant cavity simply by
detecting
photon decays from the cavities. This is a rare
example of a decay mechanism playing a constructive role in
quantum information processing. The scheme is comparatively easy to
implement, requiring only the ability to
trap a single three level atom in a cavity. Moreover,
by adding one more atom to Alice's cavity, it can be converted to
a {\em reliable} state transfer protocol. This state transfer protocol
can be viewed as an {\em alternative} to designer laser pulses for
 transferring (Refs.\cite{zol3,Pel})
quantum information
into a cavity from outside.

\section{Conclusions}

Summarising, we discussed in this paper how the interaction of a system with its environment can be used to produce entanglement in a controlled way. As an example we discussed the preparation and manipulation of entangled states of atoms inside a cavity and showed how quantum communication protocols
such as teleportation can be implemented by the detection of photon decay
from cavities containing trapped ions or atoms.

{\em Acknowledgement:} We thank P.~Grangier for fruitful
discussions. This work was also supported by the EQUIP project
of the European Union, the TMR network on Cavity QED and Non-classical
Light, the A.~v.~Humboldt Foundation, the Leverhulme Trust, the
European Science Foundation, by the UK Engineering and Physical
Sciences Research Council, DGICYT Project No.
PB-98-0191 (Spain), and by the Sonderforschungs\-be\-reich
 237 ``Unordnung und gro{\ss}e Fluktuationen".

\end{document}